\documentclass[12pt]{article}

\usepackage{amsmath}
\usepackage{amssymb}
\usepackage{dsfont} 

\usepackage{graphicx}

\usepackage[dvipsnames]{xcolor}
\usepackage{color}

\newcommand{\be}{\begin{equation}}
\newcommand{\ee}{\end{equation}}
\newcommand{\bel}[1]{\begin{equation}\label{#1}}
\newcommand{\bea}{\begin{eqnarray}}
\newcommand{\eea}{\end{eqnarray}}
\newcommand{\balign}{\begin{align}}
\newcommand{\ealign}{\end{align}}
\newcommand{\ba}{\begin{array}}
\newcommand{\ea}{\end{array}}
\newcommand{\bfig}{\begin{figure}}
\newcommand{\efig}{\end{figure}}

\newcommand{\eref}[1]{(\ref{#1})}

\allowdisplaybreaks

\newcommand{\bra}[1]{\mbox{$\langle \, {#1}\, |$}}
\newcommand{\ket}[1]{\mbox{$| \, {#1}\, \rangle$}}
\newcommand{\exval}[1]{\mbox{$\langle \, {#1}\, \rangle$}}
\newcommand{\inprod}[2]{\mbox{$\langle \, {#1} \, | \, {#2} \, \rangle$}}
\newcommand{\dbra}[1]{\mbox{$\langle\langle \, {#1}\, |$}}
\newcommand{\dket}[1]{\mbox{$| \, {#1}\, \rangle\rangle$}}

\newcommand{\dinprod}[2]{\mbox{$\langle\langle \, {#1} \, | \, {#2} \, \rangle\rangle$}}

\newcommand{\rmd}{\mathrm{d}}
\newcommand{\rme}{\mathrm{e}}

\newcommand{\half}{\frac{1}{2}}

\newcommand{\Tr}{\mathop{\mathrm{Tr}}\nolimits}

\newcommand{\comm}[2]{\mbox{$[\,{#1}\,,\,{#2}\,]$}}
\newcommand{\anti}[2]{\mbox{$\left\{{#1} , {#2}\right\}$}}

\newcommand{\C}{{\mathbb C}}
\newcommand{\R}{{\mathbb R}}

\newcommand{\N}{{\mathbb N}}

\newtheorem{theo}{Theorem}[section]
\newtheorem{lmm}[theo]{Lemma}
\newtheorem{df}[theo]{Definition}
\newtheorem{prop}[theo]{Proposition}
\newtheorem{coro}[theo]{Corollary}
\newtheorem{rem}[theo]{Remark}

\newcommand{\proof}{\noindent {\it Proof: }}
\def\qed{\hfill$\Box$\par\medskip\par\relax}

\begin{document}

\title{Matrix product ansatz for non-equilibrium quantum steady states}
\author{D. Karevski$^{1}$, \and V. Popkov$^{2}$, \and G.M.~Sch\"utz$^{3}$
}

\maketitle

{\small
\noindent $^{~1}$Institut Jean Lamour, dpt. P2M, Groupe de Physique Statistique, 
Universit\'e de Lorraine, CNRS UMR 7198, B.P. 70239, F-54506 Vandoeuvre les Nancy Cedex, France
\\
\noindent Email: dragi.karevski@univ-lorraine.fr

\smallskip
\noindent $^{~3}$Helmholtz-Institut f\"ur Strahlen-und Kernphysik, 
Universit\"at Bonn, Nussallee 14-16, 53119 Bonn, Germany
\\
\noindent Email: popkov@uni-bonn.de

\smallskip
\noindent $^{~2}$Institute of Complex Systems II,
Forschungszentrum J\"ulich, 52425 J\"ulich, Germany
\\
\noindent Email: g.schuetz@fz-juelich.de
}

\abstract{We present a general construction 
of matrix product states for stationary density matrices of 
one-dimensional quantum 
spin systems kept out of equilibrium through 
boundary Lindblad dynamics. 
As an application we review the isotropic Heisenberg quantum spin 
chain which is closely related to the generator of the simple symmetric
exclusion process. Exact and heuristic results as well as
numerical evidence suggest a local quantum equilibrium and long-range 
correlations  
reminiscent of similar large-scale properties in classical
stochastic interacting particle systems that can be understood in terms
of fluctuating hydrodynamics.
}

\section{The Quantum master equation}
\label{sec:1}

This article is concerned with stationary states of non-equilibrium
quantum spin systems, addressing a mathematically minded readership. 
We spent some effort on
recalling -- in mathematical terms -- relevant basic quantum mechanical notions
as well as providing motivations from physics as to why quantum spin systems are of
great current interest. Among them is, we feel, a striking analogy 
with non-local properties of {\it classical} stochastic interacting
particle systems \cite{Spoh83,Bert02,Derr02} that we point out 
in the hope of stimulating further mathematically rigorous 
work.

Let $\mathfrak{H}$ be a separable complex Hilbert space. A concrete physical 
quantum system is mathematically
defined by a specific self-adjoint (not necessarily bounded) linear operator $H$ 
on $\mathfrak{H}$, called quantum Hamiltonian (in the following simply Hamiltonian).
Vectors in $\mathfrak{H}$ are denoted by the ket-symbol $\ket{\cdot}$ and
vectors in the dual space $\mathfrak{H}^\ast$ are denoted by the bra-symbol
$\bra{\cdot}$. The scalar product of two vectors $\ket{\Psi} = \sum_n c_n \ket{n}
\in \mathfrak{H}$ 
and $\ket{\Phi} = \sum_n b_n \ket{n} \in \mathfrak{H}$ with coordinates 
$b_n,c_n \in \C$ in some orthonormal basis $\ket{n}$, $\bra{n}$ of 
$\mathfrak{H}$ and its dual
resp. is denoted 
$\inprod{\Phi}{\Psi}$ and defined to be linear in the {\it second} argument, i.e.,
$\inprod{\Phi}{\Psi} := \sum_n \bar{b}_n c_n$ where
the bar denotes complex conjugation.
We denote the unit operator on $\mathfrak{H}$ by $\mathbf{1}$. 
The Kronecker symbol $\delta_{a,b}$ is defined by $\delta_{a,b} = 1$
if $a=b$ and $\delta_{a,b} = 0$ else for $a$ and $b$ from any set. 

The eigenvalues $E_n$ of the Hamiltonian $H$ are the physical energies measured 
in an experiment when the physical system is in an  eigenstate $n$ of $H$, 
defined by the corresponding
eigenvector $\ket{\Psi_n}$. One normalizes these eigenvectors,
which span the Hilbert space $\mathfrak{H}$,
to satisfy the orthogonality relation 
$\inprod{\Psi_n}{\Psi_m} = \delta_{n,m}$. 
A spectral ray $\ket{\Psi} \in \mathfrak{H}$ 
normalized such that $||\Psi||^2:= \inprod{\Psi}{\Psi} = 1$
(i.e. a vector defined up to an arbitrary phase)
is called a state vector. It represents the full information that one can have about
a quantum system under the idealizing assumption that it is isolated
(and has always been isolated) from its physical environment.\footnote{Due 
to the quantum mechanical phenomenon of entanglement,
a quantum subsystem that has interacted with its environment 
in the past (until some time $t_0$)
cannot be considered isolated for $t \geq t_0$ 
even when there are no interactions from $t_0$ onwards.}
The modulus $|\psi_n|^2$ of the components of $\ket{\Psi}$ 
are the probabilities to find the physical system in eigenstate $n$. 

In general, physically observable properties of a quantum system (e.g. particle 
positions, momenta and so on) are represented by self-adjoint linear 
operators $O_i$ on $\mathfrak{H}$ which we call {\it observables}. The ``fuzzy'' 
and non-deterministic nature of quantum mechanics is reflected by the fact that 
the $O_i$ are not all diagonal in 
some fixed basis of $\mathfrak{H}$ and that only the mean outcome of a large number (mathematically 
speaking, an infinite number) of measurements of such an observable is predictable.
By the mean (or expected) value of an
observable $O$ in a general state vector $\ket{\Psi}$ we mean the scalar product 
$\exval{O} \equiv \bra{\Psi}O\ket{\Psi} 
= \sum_{m,n} \bar{c}_m c_n \bra{\Psi_m} O \ket{\Psi_n}$.

A self-adjoint positive definite linear operator on $\mathfrak{H}$ with unit 
trace is called a {\it density matrix} or {\it state} (not eigenstate !) of a physical system. Therefore a density matrix $\rho$
with eigenvalues $\rho_n \in \R$ has the properties
\bel{rhoprop}
\rho^\dagger = \rho, \quad \rho_n \geq 0, \quad \Tr(\rho) = 1
\ee
where the dagger-symbol $\dagger$ denotes hermitian conjugation. For a given Hilbert space
we denote the set of all density matrices by $\mathfrak{S}(\mathfrak{H})$.
The mean value of an observable $O_i$ in a state
$\rho$ is given by the Frobenius scalar product $\exval{O_i} := \Tr (O_i^\dagger\rho)$.

Unlike a state vector describing a {\it single and isolated} quantum system, 
a density matrix contains the full information about a quantum
system in either of the following three scenarios: 

(1) A density matrix of the specific form 
\bel{rhopure}
\rho = \ket{\Psi}\bra{\Psi}
\ee 
may describe a single isolated system.\footnote{Following quantum mechanical convention we use the short hand
$\ket{\cdot}\bra{\cdot} \equiv \ket{\cdot}\otimes\bra{\cdot}$ for the Kronecker
product $\otimes$ of a state vector $\ket{\cdot} \in \mathfrak{H}$ and some 
dual state vector
$\bra{\cdot} \in \mathfrak{H}^\ast$. We stress that by the rules of
tensor calculus one has $\bra{\Psi}\otimes\ket{\Phi} = \ket{\Psi}\otimes\bra{\Phi} 
\equiv \ket{\Psi} \bra{\Phi}$
but $\bra{\Psi}\otimes\ket{\Phi} \neq \inprod{\Psi}{\Phi}$
since $\inprod{\Psi}{\Phi}$ represents the scalar product.}
In this case we say that $\rho$ is a {\it pure} state.
If a density matrix is not a pure state then there is no state vector 
$\ket{\Psi}$
such that $\Tr (O_i^\dagger\rho) =\bra{\Psi}O\ket{\Psi}$ 
for all observables $O_i$.

(2) One describes an {\it ensemble} of identical isolated quantum systems. 
In particular, if for some $\beta \in \R^+_0$ the density matrix
is of the form
\bel{rhoeq}
\rho = \frac{1}{Z}\rme^{-\beta H}
\ee 
where $Z = \Tr\left(\exp{(-\beta H)}\right)$
we say that the physical system defined by the Hamiltonian $H$ is in
thermal equilibrium at temperature $T = 1/\beta$ and the normalization factor
$Z$ is called the partition function. In this case the probability
to find the system in an eigenstate $n$ of $H$ is proportional to the 
Boltzmann weight $\exp{(-\beta E_n)}$ analogous to classical thermodynamics.

(3) $\rho$ describes a subsystem (or an
ensemble thereof) of a larger physical system with which {\it it interacts} 
(or has interacted in the past).\footnote{For this scenario, which we have in mind for
applications, one often calls $\rho$ the 
{\it reduced density matrix}, but we shall refrain doing so here.}

Pure states and equilibrium states have in common that they remain so
when the physical system is isolated from its environment or becomes isolated from
some time $t\geq t_0$ onwards. This follows from the time-evolution
equation for the density matrix $\rho_t$ of an isolated quantum system 
with quantum Hamiltonian $H$
\bel{evouni} 
\frac{\rmd}{\rmd t} \rho_t = - i \comm{H}{\rho_t} 
\ee
where the commutator is defined by $[A,B]:= AB - BA$. Therefore
an equilibrium state is stationary. A pure state 
$\rho_0 = \ket{\Psi(0)}\bra{\Psi(0)}$ is only stationary
if $\ket{\Psi(0)}$ is an eigenstate of $H$, but generally remains a pure
state since the evolution
equation \eref{evouni} is solved by the unitary transformation 
$\rho_t = \exp{(-iHt)} \rho_0 \exp{(iHt)}$ which
gives $\rho_t = \ket{\Psi(t)}\bra{\Psi(t)}$ with 
$\ket{\Psi(t)} = \exp{(-iHt)} \ket{\Psi(0)}$. 

We are interested in 
open systems that are in contact with an environment.
In the Markovian approach to open quantum systems \cite{Atta06,Breu02} the time
evolution 
\bel{semigroup}
\rho_t = \Lambda_t \rho_0
\ee
is given by a one-parameter semigroup $\Lambda_t$ of linear endomorphisms 
on the space $\mathfrak{S}(\mathfrak{H})$ of
all density matrices \cite{Koss72}. Under some continuity conditions and for
bounded $H$ the Lindblad theorem \cite{Lind76} asserts that the 
infinitesimal generator $\mathcal{L}$ of the semigroup $\Lambda_t$
that preserves self-adjointness, positivity and unit trace is of the form
\bel{Lindblad1}
\mathcal{L}(\rho) = - i \comm{H}{\rho} + \mathcal{D} (\rho).
\ee
The commutator
describes the unitary part of the time evolution
(as in an isolated quantum system) and the dissipative part 
$\mathcal{D}(\rho)  \in  \mathfrak{End}(\mathfrak{S}(\mathfrak{H}))$, 
which encodes
the physical properties of the coupling to the environment, is of the form
\bel{Lindblad2}
\mathcal{D}(\rho) = \sum_j \mathcal{D}_j (\rho), \quad
\mathcal{D}_j (\rho) = D_j \rho D_j^\dagger 
- \half \{  \rho  , D_j^\dagger D_j \}
\ee
with 
bounded operators $D_j \in \mathfrak{End}(\mathfrak{H})$ and the anticommutator
$\anti{A}{B} := AB + BA$. The evolution equation \eref{Lindblad1} 
with dissipators \eref{Lindblad2} is called
{\it quantum master equation}.
The operators $D_j$ 
that specify an individual dissipator are called Lindblad operators.
In an open system a state that is initially pure or in equilibrium 
does not in general remain so
as would be the case in the absence of dissipators in \eref{Lindblad1}.
This raises the question of stationary states in open systems.

In order to address existence we introduce the adjoint generator 
$\mathcal{L}^\dagger$ which is defined as follows \cite{Koss72}.
Consider the Banach space $L^1(\mathfrak{H})$
over $\R$ of self-adjoint trace class linear operators $\sigma \in \mathfrak{H}$ 
with norm given by $|| \sigma ||_1 = \sup \sum_n |(x_n, \sigma y_n)|$ where the
supremum is taken over all orthonormal and complete bases $\{x_n\}$ and
$\{y_n\}$ of $\mathfrak{H}$. Then all linear, real and continuous
functionals $F$ on $L^1(\mathfrak{H})$ are of the form 
$\langle F, \sigma \rangle =
\Tr(F^\dagger \sigma)$ where $F$ is a bounded self-adjoint 
linear operator on $\mathfrak{H}$.
The set of all such bounded observables $F$ defines the space 
$L^\infty(\mathfrak{H})$ dual to $L^1(\mathfrak{H})$.
Its norm is given by $||F||_\infty = \sup_{|| \sigma ||_1 = 1} 
|\langle F, \sigma \rangle| = \sup_{\Psi \in \mathfrak{H}} ||F \Psi||/||\Psi||$. Then the adjoint
generator is given by
\bel{Lindad}
\mathcal{L}^\dagger (F) = - i [H,F] + \sum_j \left(
 D_j^\dagger F D_j 
- \half \{ F  , D_j^\dagger D_j \} \right) 
\ee
and one sees that $\mathcal{L}^\dagger (\mathbf{1}) = 0$. 
If $\mathfrak{H}$ is finite-dimensional then this guarantees the existence
of a density matrix $\rho$ such that
\bel{Lindstat}
\mathcal{L} (\rho) = 0.
\ee
We call a density matrix satisfying \eref{Lindstat} a {\it stationary state},
and, in particular, when $\rho \neq
\rme^{-\beta H}/Z$ for any $\beta \in \R^+_0$, we call $\rho$ a {\it non-equilibrium steady state}
(NESS) of the open quantum system with Hamiltonian $H$. For ergodicity and approach
to stationarity, which
are not our concern, we refer to \cite{Frig78}.
For Lindblad operators of the form $D_j = \Gamma L_j$ with a common 
coupling constant $\Gamma$
the strong 
coupling limit
$\Gamma \to \infty$ is called the {\it Zeno limit}.

Finally we remark that shifting the Lindblad operators by (in general complex)
constants $c_j$ generates an additional unitary term in the quantum master
equation. More precisely, defining for some $c_j\in\C$ the self-adjoint operators
\bel{G}
G_j = \frac{i}{2} \left(c_j D_j^\dagger - \bar{c}_j D_j\right), \quad 
\tilde{H} = H - \sum_j G_j,
\ee
one has
\bel{genshift}
\mathcal{L}(\rho) = \tilde{\mathcal{L}} (\rho)
\ee
where $\tilde{\mathcal{L}}$ is defined by the modified Hamiltonian $\tilde{H}$
and shifted Lindblad operators
\bel{Lshift}
\tilde{D}_j := D_j - c_j .
\ee
Notice that $\tilde{G}_j = G_j$.

This paper deals with the construction of non-equilibrium stationary states
$\rho$ defined by \eref{Lindstat}
for a specific family of physical systems of great interest, viz. 
quantum spin chains coupled to environment at their boundaries, 
defined in Sec.~2. 
In Sec.~3 we generalize in mathematically rigorous form the matrix product ansatz (MPA) 
of Prosen \cite{Pros11,Pros15} with local divergence condition introduced by 
us in \cite{Kare13a}. 
As an application (Sec.~4) we
summarize recent progress that we made for the 
stationary non-equilibrium magnetization profiles in the isotropic spin-1/2
Heisenberg quantum spin chain \cite{Kare13a,Kare13b,Popk16}
and discuss it in the light of very recent results \cite{Buca16} 
on correlation functions for this quantum system.
The upshot
is that there are substantial and perhaps somewhat unexpected similarities between
quantum and classical stationary states of boundary-driven non-equilibrium
systems.

\section{Quantum spin chains}
\label{sec:XXX}

\subsection{Why quantum spin chains?}

The prototypical model for the quantum mechanical description of magnetism in 
linear chains of atoms is the so-called Heisenberg 
quantum spin chain, proposed first in 1928 \cite{Heis28} as an improvement
over the classical Ising model which was introduced a few years earlier by 
Lenz and solved by his student Ernst Ising in 1925 \cite{Isin25}. The simplest 
version of the Heisenberg model,
the spin-1/2 chain defined below, is exactly solvable in the sense of quantum
integrability \cite{Baxt82}. Hence the equilibrium properties of the system, which
were derived in the past decades in a vast body of literature, are rather well 
understood from a theoretical perspective and to some extent also experimentally
for various spin-chain materials which exhibit quasi one-dimensional interactions
between neighbouring atoms.

In recent years, novel experimental Laser techniques involving single cold atoms 
in optical traps have made the investigation of spin chains {\it far from thermal 
equilibrium} feasible. The unique possibilities that the study of individual 
interacting atoms offers has triggered an immense experimental research activity. 
On the theoretical side, however, not much is known about non-equilibrium 
steady states of spin chains which are of particular interest in the case of {\it boundary
driving}, since in this way one obtains information about anomalous transport
properties. By boundary driving we mean a scenario where the two ends of a chain 
are forced into different states by some boundary interaction with the
physical environment of the chain, thus inducing stationary currents of
locally conserved quantities along the chain. The bulk of the system is considered
to be effectively isolated from its physical environment, i.e., described by some
quantum Hamiltonian $H$. The boundary interaction is described by Lindblad
dissipators.

Exact results are scarce for chains
with more than just a few atoms and there are, to our
knowledge, no exact concrete results for specific quantum chains of 
arbitrary length kept far from
thermal equilibrium by some kind of Lindblad boundary-drive. This state of
affairs is in stark contrast to
classical stochastic interacting particle systems whose Markov generators
can be expressed in terms of (non-Hermitian) quantum spin chains \cite{Schu01}
and for which many exact 
and rigorous results exist 
\cite{Derr98,Kipn99,Ligg99,Schu01,Blyt07} and which are
also amenable to generally applicable analytical approaches 
such as macroscopic fluctuation
theory \cite{Bert15} and non-linear fluctuating hydrodynamics \cite{Spoh14}.

Nevertheless, a breakthrough in the study of quantum systems far from 
thermal equilibrium came a few years ago
through the work of Prosen \cite{Pros11} who devised a matrix product ansatz (MPA) somewhat
reminiscent of the matrix product ansatz for classical 
stochastic interacting particle systems \cite{Blyt07}. This MPA was subsequently
developed by us, using a local divergence technique that reveals a
link to quantum integrability and symmetries of the quantum system \cite{Kare13a}.
The MPA allowed for the derivation of
recursion relations for mean values of physical observables from which stationary 
currents and magnetization profiles could be computed numerically exactly
for {\it large} finite chains \cite{Kare13b} and analytically from a 
continuum approximation to these
recursion relations. Generalizing the continuum approximation, also correlations 
have been obtained analytically for the Heisenberg chain \cite{Buca16}. 
As pointed out below,
these results point to an interesting analogy 
with a well-known result in classical stochastic interacting
particle systems \cite{Spoh83,Bert02,Derr02}.

\subsection{Definitions and notation}
\label{Sec:Definotat1}

The set
of integers $\{0,\dots,n-1\}$ is denoted $\S_n$.
We denote the canonical basis vectors of the $n$-dimensional complex vector space $\C^n$ 
by the symbol $|\alpha)$ with $\alpha\in\S_n$. 
Complex conjugation of some $z\in\C$ is denoted by $\bar{z}$. 
The canonical basis vectors of the dual space are denoted by $(\alpha|$.
With the scalar product 
$(w|v) := \sum_{\alpha} \bar{w}_\alpha  v_\alpha$ and norm 
$||v|| = \sqrt{\sum_{\alpha} |v_\alpha|^2}$ the vector space $\C^{n}$
becomes a 
finite-dimensional Hilbert space which we shall call the {\it local physical space}
and denote by $\mathfrak{p}$.

From the canonical basis vectors of  $\C^n$ we construct the canonical basis of the space 
$\mathfrak{End}(\C^n)$ of endomorphisms $\C^n \to \C^n$ by the Kronecker products
$E^{\alpha\beta} := |\alpha)(\beta| \equiv |\alpha)\otimes(\beta|$.
Generally we shall somewhat loosely
identify endomorphisms on some vector space with their
matrix representation and sometimes call them operators.
The $n$-dimensional matrices
$E^{\alpha\beta}$ have matrix elements $(E^{\alpha\beta})_{jk} = \delta_{\alpha,j}
\delta_{\beta,k}$ and they satisfy
\bea
\label{Ealgebra} 
E^{\alpha\beta} E^{\gamma\delta} & = & \delta_{\beta,\gamma} E^{\alpha\delta} \\
\label{Etrace}
\Tr (E^{\alpha\beta}) & = & \delta_{\alpha,\beta}.
\eea
The $n$-dimensional unit matrix is denoted by $\mathds{1}$. If a 
complex number appears as one term in any equation
for matrices, then this complex 
number is understood to be a multiple of the unit matrix.

We construct a canonical basis of $\C^{n^N}$ 
by the tensor product $\ket{\vec{a}} = |\alpha_1) \otimes \dots \otimes |\alpha_N)$
with the $N$-tuple
$\vec{a} = (\alpha_1, \dots , \alpha_N) \in \S_n^N$. A general vector
in $\C^{n^N}$ with components $v_{\vec{a}}$ is then denoted by $\ket{v}$.
We also define basis vectors
$\bra{\vec{a}}$ of the dual space
$\C^{{2^N}\ast}$ (isomorphic to $\C^{{2^N}}$) and the scalar product 
$\inprod{w}{v} := \sum_{\vec{a}\in\S^N} \bar{w}_{\vec{a}} v_{\vec{a}}$ and norm 
$||v|| = \sqrt{\sum_{\vec{a}} |v_{\vec{a}}|^2}$. With these definitions $\C^{n^N}$
becomes a 
finite-dimensional Hilbert space which we shall call the {\it physical space}
and denote by $\mathfrak{P}$.
Here and below
\be 
\sum_{\vec{a}} := \sum_{\alpha_1 \in \S}  \dots 
\sum_{\alpha_N \in \S} 
\ee
is the $N$-fold sum over all indices in $\S$. 

From arbitrary matrices $Q \in \mathfrak{End}(\C^n)$ we construct the local tensor operators 
\bel{localtensor}
Q_k = \mathds{1}^{\otimes (k-1)} \otimes Q \otimes 
\mathds{1}^{\otimes (N-k)} \in \mathfrak{End}(\mathfrak{P}).
\ee
By convention $Q^{\otimes 0} := 1$ and $Q^{\otimes 1} := Q$ for any matrix $Q$. 
We denote the unit matrix
acting on $\mathfrak{P}$ by $\mathbf{1}$, i.e., $\mathbf{1}=\mathds{1}^{\otimes N}$.
The set of products 
\bel{basisendP}
\{ E^{\vec{a},\vec{a}'}\} = \{\prod_{j=1}^N E^{\alpha_j\alpha_j'}_j \}
\ee 
for 
$\vec{a}, \vec{a}' \in \S_n^N$
forms a complete basis of $\mathfrak{End}(\mathfrak{P})$.
Transposition
of a matrix $A$ is denoted by $A^T$. The adjoint of an operator is denoted
$A^\dagger$ which in matrix form means $A^\dagger= \bar{A}^T$. 
Self-adjoint operators are called Hermitian. It is convenient to represent 
ket-vectors $\ket{v}$ as column vectors with components $v_{\vec{a}}$. 
Then $\bra{v}$ is represented by a row vector with components $\bar{v}_{\vec{a}}$.
Elements of a generic vector space $\mathfrak{V}$ (not Hilbert) over $\C$ are denoted 
by the double-ket symbol $\dket{\cdot}$ and elements of its dual $\mathfrak{V}^\ast$
by the double-bra symbol $\dbra{\cdot}$. A linear form $\phi_W: \mathfrak{V} \to \C$ 
is denoted by $\dinprod{W}{\cdot}$.

%
With these conventions we are now in a position to define the objects of our
investigation.

\begin{df}
\label{Def:QH}
Let $h \in \mathfrak{End}(\C^{n^2})$  and $b^L,b^R \in \mathfrak{End}(\C^n)$ 
be self-adjoint
and $b^L_1 = b^L \otimes \mathds{1}^{\otimes (N-1)}$, 
$b^R_N = \mathds{1}^{\otimes (N-1)} \otimes b^R$,
$h_{k,k+1} = \mathds{1}^{\otimes (k-1)} \otimes h \otimes \mathds{1}^{\otimes (N-k-1)}$.
Then a homogeneous quantum spin chain with $N \geq 2$ sites
with nearest-neighbour interaction $h$ and boundary fields $b^{L,R}$
is defined by the Hamiltonian
\bel{QHdef}
H = b^L_1 + b^R_N + \sum_{k=1}^{N-1} h_{k,k+1} .
\ee
A quantum spin system with one site is defined by a self-adjoint operator 
$b\in \mathfrak{End}(\C^n)$.
\end{df}

\begin{df}
\label{Def:DissQH}
For $D^{\chi_k} \in \mathfrak{End}(\C^n)$ and a density matrix 
$\rho \in \mathfrak{S}(\mathfrak{P})$ the operator 
\be 
\mathcal{D}_k(\rho) := D^{\chi_k}_k\rho D_k^{\chi_k\dagger} 
- \half \left(\rho D_k^{\chi_k\dagger} D^{\chi_k}_k
+ D_k^{\chi_k\dagger} D^{\chi_k}_k\rho\right), \quad 1 \leq k \leq N, \quad N \geq 1
\ee
is called dissipator at site $k$ with local Lindblad operator $D^{\chi_k}$, indexed
by a symbol $\chi_k$. 
For $N=1$ the lower index $k=1$ is dropped.
\end{df}

\begin{df}
\label{Def:NESSQH}
Let $H$ be a quantum spin Hamiltonian with $N$ sites according to Definition \ref{Def:QH},
$\mathcal{D}_1$ and $\mathcal{D}_N$ be dissipators with local Lindblad operators 
$D^L$ and $D^R$ resp. according to Definition \ref{Def:DissQH}
and let $\rho \in \mathfrak{S}(\mathfrak{P})$ be the solution of the equation
\bel{NESSQHdef}
-i \comm{H}{\rho} + \mathcal{D}_1(\rho) + \mathcal{D}_N(\rho) = 0.
\ee
Then $\rho$ is called a non-equilibrium stationary state of the boundary-driven
quantum spin system defined by $H$.
\end{df}

We remark that the construction of matrix product states given below is straightforwardly
generalized to more than one boundary dissipator at each edge of the chain.

\section{Construction of stationary matrix product states}

\subsection{Matrix product ansatz}

In order to construct a solution of the stationary Lindblad equation of the form
\eref{NESSQHdef} we
first make the following observations:\\ 
\noindent (a) For any density matrix $\rho \in \mathfrak{S}(\mathfrak{P})$ one can find a matrix $M \in \mathfrak{End}(\mathfrak{P})$ such that
\bel{rhodeco}
\rho = M M^\dagger / Z
\ee
with the {\it partition function}
\bel{partfundef}
Z := \Tr(M M^\dagger).
\ee
Thus, given $M$ one knows $\rho$.\footnote{$M$ is not uniquely defined.
For a given $M$ and arbitrary unitary $U$ the product $M U$ gives the same $\rho$. 
This non-uniqueness seems to be exactly the point that makes $M$ easier to treat 
than $\rho$.}\\
\noindent (b) One can expand $M$ in the basis \eref{basisendP}
of $\mathfrak{End}(\mathfrak{P})$ as
\bel{Sdeco}
M = \sum_{\vec{a}, \vec{a}'} 
M_{\vec{a}, \vec{a}'} 
E^{\alpha_1,\alpha_1'}_1\dots E_N^{\alpha_N,\alpha_N'}.
\ee

The idea of the matrix product ansatz (MPA) 
is to write the matrix elements $M_{\vec{a}, \vec{a}'}$ as the linear form 
\cite{Pros11,Pros15}
\bel{SMPA}
M_{\vec{a}, \vec{a}'}  = \dbra{W} 
\Omega^{\alpha_1,\alpha_1'}\dots\Omega^{\alpha_N,\alpha_N'} \dket{V}
\ee
where $\dket{V}$ is a vector in some (generally infinite-dimensional) auxiliary space 
$\mathfrak{A}$, the $n^2$ matrices $\Omega^{\alpha,\alpha'}$ are suitably chosen endomorphisms of $\mathfrak{A}$
and $\dbra{W}$ is a suitably
chosen vector from the dual space $\mathfrak{A}^\ast$.

In order to use this MPA in applications we need to add some more structure.
We define $\bar{\Omega}^{\alpha,\alpha'} \in \mathfrak{End}(\mathfrak{A})$ 
by complex conjugation
of the matrix representation of $\Omega^{\alpha,\alpha'}$. Next we construct
\bea
\label{Omegadef}
\Omega & := & \sum_{\alpha,\alpha'} E^{\alpha\alpha'} \otimes \Omega^{\alpha\alpha'}, \quad
\Omega^\star :=  \sum_{\alpha,\alpha'} E^{\alpha\alpha'} \otimes \bar{\Omega}^{\alpha'\alpha} 
\quad \in \mathfrak{End}(\C^n\otimes\mathfrak{A})\\
\label{Omegatensdef}
\Omega^{\otimes_p N} & := & \sum_{\vec{a}, 
\vec{a}'} E^{\alpha_1\alpha_1'} \otimes \dots \otimes E^{\alpha_N\alpha_N'} \otimes \Omega^{\alpha_1\alpha_1'}\dots\Omega^{\alpha_N\alpha_N'}
\in \mathfrak{End}(\mathfrak{P}\otimes\mathfrak{A}) 
\eea
and analogously $\left(\Omega^\star\right)^{\otimes_p N} 
= \left(\Omega^{\otimes_p N}\right)^\star$. The subscript $p$ at the tensor symbol
indicates that the tensor product is only taken over the local physical space 
$\mathfrak{p}$, i.e., the term 
$\Omega^{\alpha_1\alpha_1'}\dots\Omega^{\alpha_N\alpha_N'} \in 
\mathfrak{End}(\mathfrak{A})$ in \eref{Omegatensdef} is the usual matrix product.
The star $\star$
denotes the adjoint operation on the physical space 
$\mathfrak{P}$ only, 
not on the auxiliary space. 
This means that
the matrix $\left(\Omega^{\otimes_p N}\right)^\star$
is obtained from the matrix
$\Omega^{\otimes_p N}$ by transposition and complex conjugation of 
its components $\Omega^{\otimes_p N}_{\vec{a}, 
\vec{a}'} =
\Omega^{\alpha_1\alpha_1'}\dots\Omega^{\alpha_N \alpha_N'}
\mapsto \bar{\Omega}^{\alpha_1'\alpha_1}\dots\bar{\Omega}^{\alpha_N' \alpha_N}
=\left(\Omega^{\otimes_p N}\right)^\star_{\vec{a}, 
\vec{a}'}$
as in the second definition in \eref{Omegadef}
{\it without} reversing the order of the matrix products and 
{\it without} transposing the matrices $\Omega^{\alpha_j\alpha_j'}$.

This construction immediately leads to the following lemma:

\begin{lmm}
\label{Mrep}
Let $\mathfrak{A}$ be a vector space, $\Omega^{\alpha,\alpha'} \in \mathfrak{End}
(\mathfrak{A})$ for $\alpha,\alpha'\in \S_n$, $\dket{V},\dket{\overline{V}}\in\mathfrak{A}$ 
and $\dbra{W},\dbra{\overline{W}}\in\mathfrak{A}^\ast$ where the bar denotes complex
conjugation of each vector component.
Then $M, M^\dagger \in \mathfrak{End}(\mathfrak{P})$ defined by \eref{Sdeco} and
\eref{SMPA} can be written
\bel{Mtensor}
M = \dbra{W} \Omega^{\otimes_p N} \dket{V}, \quad 
M^\dagger = \dbra{\overline{W}} \left(\Omega^\star\right)^{\otimes_p N} \dket{\overline{V}}
\ee
where
the  linear form $\dbra{W} \cdot \dket{V}$ 
on $\mathfrak{A}$ is taken on each component 
$\Omega^{\alpha_1\alpha_1'}\dots\Omega^{\alpha_N\alpha_N'}$ of the
endomorphism
$\Omega^{\otimes_p N}$ on $\mathfrak{P} \otimes \mathfrak{A}$.
\end{lmm}

Lemma \eref{Mrep} follows immediately from the expansion $\Omega = \sum_{\alpha,\alpha'} E^{\alpha\alpha'} \otimes \Omega^{\alpha\alpha'}$ and the multilinearity of the Kronecker
product. It expresses the fact that $\Omega^{\otimes N}$ can be thought of as a matrix
of dimension $n^N$ (the dimension of the physical space $\mathfrak{P}$) whose
matrix elements $(\vec{a},\vec{a}')$ are the products 
$\Omega^{\alpha_1\alpha_1'}\dots\Omega^{\alpha_N\alpha_N'}$ of 
(generally infinite-dimensional) matrices acting on the auxiliary space 
$\mathfrak{A}$. The linear form $\dbra{W} \cdot \dket{V}$ maps each of these matrix
products
onto $\C$ so that $M$ is indeed a usual matrix of dimension $n^N$.

The next technical idea is to double the auxiliary space. 
To this end we denote the unit operator on $\mathfrak{A}$ by $I$
and define $\Omega^{\alpha\alpha'}_1 := \Omega^{\alpha\alpha'} \otimes I$
and $\Omega^{\alpha\alpha'}_2 := I \otimes \Omega^{\alpha\alpha'}$ which are endomorphisms
of $\mathfrak{A}^2$. The multilinearity of the tensor product
allows us to write $\Omega^{\alpha\alpha'}_1 \Omega^{\beta\beta'}_2 = \Omega^{\alpha\alpha'} \otimes \Omega^{\beta\beta'}$
for any $\alpha,\alpha',\beta,\beta' \in \S$. We also define in analogy to \eref{Omegadef}
the following endomorphisms of $\mathfrak{P}\otimes\mathfrak{A}^2$
\bea
\label{Omegadef3}
& & \Omega_1 := \sum_{\alpha,\alpha'}  
E^{\alpha\alpha'} \otimes \Omega^{\alpha\alpha'} \otimes I, \quad 
\Omega_2 := \sum_{\alpha,\alpha'} 
E^{\alpha\alpha'} \otimes I \otimes \Omega^{\alpha\alpha'} \\
\label{Omegadef4}
& & \Omega_1^\star := \sum_{\alpha,\alpha'}  
E^{\alpha\alpha'} \otimes \bar{\Omega}^{\alpha'\alpha} \otimes I,
\quad
\Omega_2^\star := \sum_{\alpha,\alpha'} 
E^{\alpha\alpha'} \otimes I \otimes \bar{\Omega}^{\alpha'\alpha}.
\eea

\begin{lmm}
\label{rholemma}
Let $\mathfrak{A}$ be a vector space 
and $\dket{V},\dket{\overline{V}}\in\mathfrak{A}$ and $\Omega^{\alpha\alpha'} \in \mathfrak{End}(\mathfrak{A})$ for $\alpha,\alpha' \in \S_n$ and
$\dbra{W},\dbra{\overline{W}}\in\mathfrak{A}^\ast$.
For some $\Theta^{\alpha\alpha'} \in \mathfrak{End}(\mathfrak{A}^2)$ define 
\bea
\label{Thetadef}
\Theta & := & \sum_{\alpha,\alpha'} E^{\alpha\alpha'} \otimes \Theta^{\alpha\alpha'} 
\in \mathfrak{End}(\C^2\otimes\mathfrak{A}^2) \\
\Theta^{\otimes_p N} & := & \sum_{\vec{a},\vec{a}'} E^{\alpha_1\alpha_1'} \otimes \dots \otimes E^{\alpha_N\alpha_N'} \otimes \Theta^{\alpha_1\alpha_1'}\dots\Theta^{\alpha_N\alpha_N'}
\in \mathfrak{End}(\mathfrak{P}\otimes\mathfrak{A}^2).
\eea
Then for
\bel{Theta} 
\Theta = \Omega_1 \Omega_2^\star
\ee
a density matrix $\rho \in \mathfrak{S}(\mathfrak{P})$ has the matrix product 
representation
\bel{rhoxi}
\rho =  \dbra{W,\overline{W}} \Theta^{\otimes_p N} \dket{V,\overline{V}}.
\ee
where the tensor products
\bel{scalardouble}
\dket{V,\overline{V}} := \dket{V}\otimes\dket{\overline{V}} \in \mathfrak{A} \otimes 
\mathfrak{A}, \quad 
\dbra{W,\overline{W}} := \dbra{W}\otimes\dbra{\overline{W}} \in \mathfrak{A}^\ast \otimes 
\mathfrak{A}^\ast
\ee
define a bilinear form $\phi_{W,\overline{W}}: \mathfrak{A} \otimes \mathfrak{A} \to \C$.
\end{lmm}

\noindent {\it Proof:} We first note that for the scalar product on the physical
space $\mathfrak{P}$ we have
\be 
\bra{\vec{a}} \Theta^{\otimes_p N} \ket{\vec{a}'} =
\Theta^{\alpha_1\alpha_1'}\dots\Theta^{\alpha_N\alpha_N'}.
\ee
Furthermore, by the construction \eref{Theta} for a single site and definition
\eref{Omegadef3}, one finds
\bel{Thetaab} 
\Theta^{\alpha\alpha'} = \sum_\beta (\alpha|\Omega_1 |\beta)(\beta|\Omega_2^\star |\alpha') 
=  \sum_\beta \Omega_1^{\alpha\beta} (\Omega_2^\star)^{\beta\alpha'} 
= \sum_\beta \Omega_1^{\alpha\beta} \bar{\Omega}_2^{\alpha'\beta}
\ee
and therefore with $\vec{b} := (\beta_1,\dots,\beta_N) \in \S_n^N$
\bea
\Theta^{\alpha_1\alpha_1'}\dots\Theta^{\alpha_N\alpha_N'} & = & \sum_{\vec{b}} \Omega_1^{\alpha_1\beta_1} \bar{\Omega}_2^{\alpha_1'\beta_1} \dots \Omega_1^{\alpha_N\beta_N} \bar{\Omega}_2^{\alpha_N'\beta_N} \nonumber \\
\label{Thetadouble}  
& = &
\sum_{\vec{b}} \Omega^{\alpha_1\beta_1} \dots \Omega^{\alpha_N\beta_N} \otimes
\bar{\Omega}^{\alpha_1'\beta_1} \dots \bar{\Omega}^{\alpha_N'\beta_N}.
\eea
This shows that $\Theta^{\alpha_1\alpha_1'}\dots\Theta^{\alpha_N\alpha_N'} 
\in \mathfrak{End}(\mathfrak{A^2})$ is decomposable into a finite sum of
endomorphisms of $\mathfrak{A} \otimes \mathfrak{A}$.
Then the factorization property of the scalar product involving
the tensor vectors \eref{scalardouble}
and the tensor operators \eref{Thetadouble} and Lemma \ref{Mrep} give
\bea 
\dbra{W,\overline{W}} \Theta^{\otimes_p N} \dket{V,\overline{V}}_{\vec{a}\vec{a}'} & = &
\dbra{W,\overline{W}} \Theta^{\alpha_1\alpha_1'}\dots\Theta^{\alpha_N\alpha_N'} \dket{V,\overline{V}} 
\nonumber \\
& = & \sum_{\vec{b}} 
\dbra{W} \Omega^{\alpha_1\beta_1} \dots \Omega^{\alpha_N\beta_N} \dket{V}
\overline{\dbra{W} \Omega^{\alpha_1'\beta_1} \dots \Omega^{\alpha_N'\beta_N} \dket{V}}
\nonumber \\
& = & \sum_{\vec{b}} 
M_{\vec{a}\vec{b}} 
\bar{M}_{\vec{a}'\vec{b}}
\eea
The l.h.s. of the first equation is the matrix element 
$\rho$.
Observing that $\bar{M}_{\vec{a}'\vec{b}} = M^\dagger_{\vec{b}\vec{a}'}$ and completeness
of the basis \eref{basisendP} shows that the
r.h.s. of the last equation is equal to $(MM^\dagger)_{\vec{a}\vec{a}'}$. Thus \eref{rhoxi} is proved for each matrix element
of $\rho$. \hfill{ } \qed

The point of this lemma is the fact that a matrix product form of $M$
induces a matrix product form for $\rho$ which allows for a computation of
physical observables in terms of the matrices $\Omega^{ss'}$. This is
the content of the following proposition.

\begin{prop}
\label{Zprop}
Let $\rho \in  \mathfrak{S}(\mathfrak{P})$ be a density matrix with
partition function $Z$ \eref{partfundef} and
$\dket{V,\overline{V}}$, $\dbra{W,\overline{W}}$ as defined in \eref{scalardouble}.
With
\bel{Theta0def}
\Theta_0 := \sum_{\alpha} \Theta^{\alpha\alpha} = \sum_{\alpha\beta} 
\Omega_1^{\alpha\beta} \bar{\Omega}_2^{\alpha\beta}  \quad \in 
\mathfrak{End}\left(\mathfrak{A}^2\right)
\ee
one has
\bea 
\label{ZMPA}
& & Z = \dbra{W,\overline{W}} \Theta_0^N \dket{V,\overline{V}}  \\
\label{ObsMPA}
& & \exval{E_{k_1}^{\alpha_1\alpha_1'} E_{k_2}^{\alpha_2\alpha_2'}\dots
E_{k_n}^{\alpha_n\alpha_n'}} \nonumber \\
& & = \dbra{W,\overline{W}} \Theta_0^{k_1-1} \Theta^{\alpha_1'\alpha_1} \Theta_0^{k_2-k_1-1} 
\Theta^{\alpha_2'\alpha_2} \dots \Theta^{\alpha_n'\alpha_n} \Theta_0^{N-k_n}\dket{V,\overline{V}} / Z
\eea
\end{prop}

\proof
The equality following the definition in \eref{Theta0def} follows from \eref{Thetaab}. 
By construction we have for the partition function \eref{ZMPA}
\bea 
Z & = & \sum_{\vec{a},\vec{a}'} 
\Tr\left(  
 E_1^{\alpha_1\alpha_1'}\dots E_N^{\alpha_N\alpha_N'} \right)
\dbra{W,\overline{W}} \Theta^{\alpha_1\alpha_1'} \dots\Theta^{\alpha_N\alpha_N'} 
\dket{V,\overline{V}} \nonumber \\
& = & \sum_{\vec{a},\vec{a}'} \left(\prod_{j=1}^N \Tr\left(  
 E^{\alpha_j\alpha_j'}\right)\right)
\dbra{W,\overline{W}} \Theta^{\alpha_1\alpha_1'} \dots\Theta^{\alpha_N\alpha_N'} 
\dket{V,\overline{V}}
\eea
where in the second equality we have used the factorization property of the trace
for tensor products. The trace property \eref{Etrace} 
yields the expression \eref{ZMPA} for the partition function $Z$. The expression
\eref{ObsMPA} follows in similar fashion by noting that due to \eref{Ealgebra} one has
$\Tr(E^{\alpha\alpha'}E^{\beta\beta'}) = \delta_{\alpha,\beta'}\delta_{\alpha',\beta}$.
\hfill {} \qed

\begin{rem}
Since an observable $O_k \in \mathfrak{End}(\mathfrak{P})$ can be expanded 
$O_k = \sum_{\alpha\alpha'} O^{\alpha\alpha'}_k E^{\alpha\alpha'}_k$ with numerical
coefficients of the form $O^{\alpha\alpha'}_k = \bar{O}^{\alpha'\alpha}_k \in \C$, 
Proposition \ref{Zprop} allows for computing averages of products of local
observables in terms of matrix products involving the matrices $\Theta^{\alpha\alpha'}$
and $\Theta_0$.
\end{rem}

We note two useful corollaries of Lemma \ref{rholemma} which 
follow directly from \eref{Thetaab}.

\begin{coro}
\label{disscoro}
Let $\rho$ be a density matrix according to Lemma \ref{rholemma} and $D_k
= \mathds{1}^{\otimes (k-1)} \otimes D \otimes \mathds{1}^{\otimes (N-k)}$ 
be a Lindblad operator acting non-trivially only on site $k$ with some local
Lindblad operator $D \in \mathfrak{End}(\C^n)$. 
Then for the local dissipator
$\mathcal{D}_k$ with Lindblad operator $D_k$ one has
\be 
\mathcal{D}_k(\rho) = \frac{1}{Z} \dbra{W,\overline{W}} \Theta^{\otimes (k-1)}
\otimes \Delta \otimes \Theta^{\otimes (N-k)} \dket{V,\overline{V}}
\ee
with $Z$ of Proposition \ref{Zprop} and
\be 
\Delta = \sum_{\beta} \sum_{\alpha\alpha'} 
\mathcal{D} \left(E^{\alpha\alpha'}\right) \otimes 
\Omega^{\alpha\beta}_1 \bar{\Omega}^{\alpha'\beta}_2
\ee
where $\mathcal{D}$ is the dissipator with the local Lindblad operator $D$.
\end{coro}

\begin{coro}
\label{bfcoro}
Let $\rho$ be a density matrix according to Lemma \ref{rholemma} and $b_k
= \mathds{1}^{\otimes (k-1)} \otimes b \otimes \mathds{1}^{\otimes (N-k)}
\in \mathfrak{End}(\mathfrak{P})$ be
a self-adjoint operator acting non-trivially only on site $k$ with some local
self-adjoint operator $b \in \mathfrak{End}(\C^n)$. 
Then for the unitary part of the time-evolution of the density matrix under $b_k$ one has
\be 
-i \comm{b_k}{\rho} = \frac{1}{Z} \dbra{W,\overline{W}} \Theta^{\otimes (k-1)}
\otimes \Gamma \otimes \Theta^{\otimes (N-k)} \dket{V,\overline{V}}
\ee
with $Z$ of Proposition \ref{Zprop} and
\be 
\Gamma = -i \sum_{\beta} \sum_{\alpha\alpha'} 
\comm{b}{E^{\alpha\alpha'}} \otimes 
\Omega^{\alpha\beta}_1 \bar{\Omega}^{\alpha'\beta}_2.
\ee
\end{coro}

For $D=\sum_{\alpha\alpha'} D_{\alpha\alpha'}E^{\alpha\alpha'}$,
$b=b^\dagger = \sum_{\alpha\alpha'} b_{\alpha\alpha'}E^{\alpha\alpha'}$
we note
\bea
\label{DissE}
\mathcal{D}(E^{\alpha\alpha'}) & = & \sum_{\beta\beta'}
\left( D_{\beta\alpha} \bar{D}_{\beta'\alpha'} E^{\beta\beta'} -
\half D_{\beta\beta'} \bar{D}_{\beta\alpha'} E^{\alpha\beta'} -
\half D_{\beta'\alpha} \bar{D}_{\beta'\beta} E^{\beta\alpha'} \right) \\
\label{commE}
\comm{b}{E^{\alpha\alpha'}} & = & \sum_{\beta}
\left( b_{\beta\alpha} E^{\beta\alpha'} - \bar{b}_{\beta\alpha'} E^{\alpha\beta} \right) 
\eea
which follows from \eref{Ealgebra} by straightforward computation and $b=b^\dagger$.

\subsection{Main result}

The previous discussion is ``abstract nonsense'' in so far as we have provided no
information about the matrices $\Omega^{\alpha\alpha'}$ and the vectors
$\dbra{W}$ and $\dket{V}$ from which a stationary density matrix $\rho$ 
solving \eref{NESSQHdef} could be
constructed. In order to state a sufficient property of the $\Omega^{\alpha\alpha'}$
we define
the local divergence condition which was first introduced for $n=2$ in 
\cite{Kare13a}.

\begin{df}
\label{Def:LDC} (Local divergence condition)
Let $H$ be a quantum spin Hamiltonian according to Definition \ref{Def:QH} 
and with finite local physical space $\mathfrak{p}$ and
let $\mathfrak{A}$ be a vector space with unit operator denoted by
$I$. For $\Omega^{\alpha\alpha'}, \Xi^{\alpha\alpha'} \in \mathfrak{End}(\mathfrak{A})$ 
define $\Omega := 
\sum_{\alpha\alpha'} E^{\alpha\alpha'} \otimes \Omega^{\alpha\alpha'} \in \mathfrak{End}(\mathfrak{p}
\otimes \mathfrak{A})$,
$\Xi := 
\sum_{\alpha\alpha'} E^{\alpha\alpha'} \otimes \Xi^{\alpha\alpha'} \in \mathfrak{End}(\mathfrak{p}
\otimes \mathfrak{A})$,
 and $\hat{h} := h \otimes I
\in \mathfrak{End}(\mathfrak{p}^2
\otimes \mathfrak{A})$. We say that $H$ satisfies a local divergence condition
w.r.t. some non-zero $\Omega$ and $\Xi$ if 
\bel{LDC}
\comm{\hat{h}}{\Omega \otimes_p \Omega} = \Xi \otimes_p \Omega - \Omega \otimes_p \Xi
\ee
where the tensor
product $\otimes_p$ over the physical space is defined by 
$\Xi \otimes_p \Omega := \sum_{\alpha\alpha'} \sum_{\beta\beta'}
E^{\alpha\alpha'} \otimes E^{\beta\beta'} \otimes (\Xi^{\alpha\alpha'} \Omega^{\beta\beta'})$.
\end{df}

\begin{rem}
The local divergence condition \eref{LDC} defines a quadratic algebra \cite{Poli05} 
for $2n^2$ generators $\Omega^{\alpha\alpha'}$ and $\Xi^{\alpha\alpha'}$. 
Quadratic algebras arise e.g. as universal enveloping algebras of
Lie algebras and also play an important role in the theory of quantum groups.
They also arise in the study of invariant measures of stochastic interacting
particle systems \cite{Alca98,Blyt07}. The local divergence condition can
be generalized to include a term $\hat{T} \Omega \otimes_p \Omega - \hat{T} \Omega \otimes_p \Omega$ where $\hat{T} = \mathds{1} \otimes \mathds{1} \otimes 
T$ and $T \in \mathfrak{End}(\mathfrak{A})$ \cite{Popk15}. This extension
gives rise to a cubic algebra.
\end{rem}

Next we define the {\it Lindblad boundary matching condition} which underlies in some 
shape or form many concrete applications of the MPA \cite{Pros15}, but which to our 
knowledge has never been stated as such and in full generality.

\begin{df}
\label{Def:BMDC} (Lindblad boundary matching condition)
Let $\mathfrak{A}$ be a vector space. For $\dket{V} \in \mathfrak{A}$ and 
$\dbra{W} \in \mathfrak{A}^\ast$ define the vectors 
$\dket{V,\overline{V}}:=\dket{V} \otimes \dket{\overline{V}}$ and
$\dbra{W,\overline{W}} := \dbra{W}\otimes\dbra{\overline{W}}$
and for $\Omega^{\alpha\alpha'}, \bar{\Omega}^{\alpha\alpha'},
\Xi^{\alpha\alpha'}, \bar{\Xi}^{\alpha\alpha'} 
\in \mathfrak{End}(\mathfrak{A})$ define the endomorphisms
\bel{Laa} 
\Lambda^{\alpha\alpha'} := i \sum_{\beta} 
\left( \Omega^{\alpha\beta} \otimes \bar{\Xi}^{\alpha'\beta} -
\Xi^{\alpha\beta} \otimes \bar{\Omega}^{\alpha'\beta} \right)
\ee
and for $B\in\{L,R\}$ with $b^{B}_{\alpha\alpha'} = \bar{b}^{B}_{\alpha'\alpha}\in \C$,
$D^{B}_{\alpha\alpha'} \in \C$
\bel{Gaa} 
\Gamma_B^{\alpha\alpha'} := -i \sum_{\beta\beta'}
\left( b^B_{\alpha\beta} 
\Omega^{\beta\beta'} \otimes \bar{\Omega}^{\alpha'\beta'} 
- \bar{b}^B_{\alpha'\beta}  
\Omega^{\alpha\beta'} \otimes \bar{\Omega}^{\beta\beta'} \right)  
\ee
\bea 
\Delta_B^{\alpha\alpha'} & := & \sum_{\beta\beta'} \sum_{\gamma}
\left( D^B_{\alpha\beta} \bar{D}^B_{\alpha'\beta'}  
\Omega^{\beta\gamma} \otimes  \bar{\Omega}^{\beta'\gamma} \right. \nonumber \\
\label{Daa}
& & \left.
- \half D^B_{\beta\alpha'} \bar{D}^B_{\beta\beta'} 
\Omega^{\alpha\gamma} \otimes \bar{\Omega}^{\beta'\gamma}
- \half D^B_{\beta'\beta} \bar{D}^B_{\beta'\alpha}  
\Omega^{\beta\gamma} \otimes  \bar{\Omega}^{\alpha'\gamma} \right).
\eea
We say that vectors $\dket{V} \in \mathfrak{A}$ and $\dbra{W} \in \mathfrak{A}^\ast$
satisfy the {\it Lindblad boundary
matching condition}
w.r.t. $\Omega$ and $\Xi$ if for all $\alpha,\alpha' \in \S_n$
\bel{LBMC}
0 = \dbra{X} \left(\Gamma_R^{\alpha\alpha'} + \Delta_R^{\alpha\alpha'} - \Lambda^{\alpha\alpha'}\right)
\dket{V,\overline{V}} = 
\dbra{W,\overline{W}} \left(\Gamma_L^{\alpha\alpha'} + \Delta_L^{\alpha\alpha'} + \Lambda^{\alpha\alpha'}\right)\dket{Y}
\ee
for all $\dbra{X} \in \mathfrak{A}^{2\ast}$ and all
$\dket{Y} \in \mathrm{span}(\Theta^{\alpha_2,\alpha_2'} \dots
\Theta^{\alpha_N,\alpha_N'} \dket{V,\overline{V}})$ for $N\geq 2$. 
\end{df}

\begin{rem}
Define $\Lambda_0 := \sum_\alpha \Lambda^{\alpha\alpha}$. It is easy to see that
$0 = \sum_\alpha \Delta_B^{\alpha\alpha} = \sum_\alpha \Gamma_B^{\alpha\alpha}$.
Hence \eref{LBMC} implies 
$0 = \dbra{X} \Lambda_0 \dket{V,\overline{V}} = \dbra{W,\overline{W}} \Lambda_0 \dket{Y}$. For the extended local divergence condition with operator $T$
the Lindblad boundary matching condition acquires an extra term 
$\{T,\Omega^{\alpha\alpha'}\}$ in both brackets in \eref{LBMC}.
\end{rem}

With these preparations we are in a position to state the main result
in terms of the original local divergence condition \eref{LDC}.
The adaptation to the extended local divergence condition is trivial.

\begin{theo}
Given a quantum spin Hamiltonian 
\be 
H = H_b + H_s
\ee 
according to Definition \ref{Def:QH} 
with bulk part $H_b = \sum_{k=1}^{N-1} h_{k,k+1}$ and surface part $H_s = b^L_1 + b^R_N$,
and given a vector space $\mathfrak{A}$, 
let $\Omega^{\alpha\alpha'}, \Xi^{\alpha\alpha'} \in \mathfrak{End}(\mathfrak{A})$ 
be representation matrices of the
quadratic algebra \eref{LDC} defined by $h$, 
and let $\ket{V}$ and $\ket{W}$ be vectors satisfying the
Lindblad boundary matching condition \eref{LBMC} with coefficients 
$L^B_{\alpha\alpha'}\in\C$
and $b^B_{\alpha\alpha'} = (\alpha|b^B|\alpha')$ for $B\in\{L,R\}$. 
Then a density matrix $\rho$ in the
matrix product form \eref{rhoxi}
is a stationary solution of the quantum master equation \eref{NESSQHdef} with 
Lindblad operators $L^B$ given by $L^B_{\alpha\alpha'}
= (\alpha|L^B|\alpha')$.
\end{theo}

This theorem breathes life into the matrix product form \eref{rhoxi} of the
stationary density matrix by providing
sufficient (but not necessary !) conditions on the matrices $\Omega^{\alpha\alpha'}$,
vectors $\dbra{W},\dket{V}$ and the auxiliary matrices $\Xi^{\alpha\alpha'}$.
The basic idea of the proof is
to split the quantum master equation into a bulk part and a boundary part. 
The bulk part comes from the
unitary part of the evolution under the action of $H_b$ and leads through 
the local divergence 
condition (\ref{LDC}) to a quadratic algebra for the matrices 
$\Omega^{\alpha\alpha'},\Xi^{\alpha\alpha'}$
plus some
boundary terms. The boundary part, which involves (i) these boundary terms, (ii) the
unitary evolution under the boundary fields, and (iii)
the Lindblad dissipators then becomes a set of equations for the vectors 
$\bra{W}$ and $\ket{V}$. Choosing a representation for the quadratic algebra
and fixing these vectors to satisfy the Lindblad boundary matching condition
then guarantees stationarity. 

\begin{proof}

We decompose $\rho=MM^\dagger/Z$ where $Z=\Tr(MM^\dagger)<\infty$ since
$\mathrm{dim}(\mathfrak{P})<\infty$. Hence it suffices to prove 
\bel{LMM} 
\mathcal{L}(MM^\dagger) 
:= -i\comm{H}{MM^\dagger} + \mathcal{D}_1 (MM^\dagger) + \mathcal{D}_N (MM^\dagger) = 0
\ee
for $M$ and $M^\dagger$ given by Lemma \ref{Mrep}.

We consider first the bulk part of the unitary evolution.
By definition of the commutator one has 
$\comm{H}{MM^\dagger} = \comm{H}{M}M^\dagger + M \comm{H}{M^\dagger}$.
The quadratic algebra \eref{LDC} ensures validity of the local divergence
condition according to Definition \ref{Def:LDC}.
The telescopic property of the sum in $H_b$ then implies
for $\hat{H}_b := H_b \otimes I \in \mathfrak{End}(\mathfrak{P}
\otimes \mathfrak{A})$ the commutation relation
\be
\comm{\hat{H}_b}{\Omega^{\otimes_p N}} = 
\Xi \otimes_p \Omega^{\otimes_p (N-1)} - \Omega^{\otimes_p (N-1)} \otimes_p \Xi.
\ee
and by transposition and complex conjugation in the physical space $\mathfrak{P}$
\be
\comm{\hat{H}_b}{(\Omega^\star)^{\otimes_p N}} = 
(\Omega^\star)^{\otimes_p (N-1)} \otimes_p \Xi^\star -
\Xi^\star \otimes_p (\Omega^\star)^{\otimes_p (N-1)} 
\ee
where
\be 
\Xi^\star = \sum_{\alpha\alpha'} E^{\alpha\alpha'} \otimes \bar{\Xi}^{\alpha'\alpha}.
\ee

Therefore, with 
\be 
N_L := \dbra{W} \Xi \otimes_p \Omega^{\otimes_p (N-1)} \dket{V}, \quad
N_R := \dbra{W} \Omega^{\otimes_p (N-1)} \otimes_p \Xi \dket{V}   
\ee
and consequently
\be 
N^\dagger_L = 
\dbra{\overline{W}} \Xi^\star \otimes_p (\Omega^\star)^{\otimes_p (N-1)} 
\dket{\overline{V}}, \quad
N^\dagger_R = 
\dbra{\overline{W}} (\Omega^\star)^{\otimes_p (N-1)} \otimes_p \Xi^\star 
\dket{\overline{V}}
\ee
one has
\be 
\comm{H_b}{M} = N_L - N_R, \quad \comm{H_b}{M^\dagger} = N^\dagger_R - N^\dagger_L.
\ee
This yields
\be 
-i \comm{H_b}{MM^\dagger} = i M(N^\dagger_L - N^\dagger_R) -i (N_L - N_R)M^\dagger .
\ee

Now notice that
\bea 
M N_L^\dagger & = & \dbra{W,\overline{W}} \Omega_1 \Xi_2^\star \otimes_p 
\Theta^{\otimes_p (N-1)} \dket{V,\overline{V}}  \\
N_L M^\dagger & = & \dbra{W,\overline{W}} \Xi_1 \Omega_2^\star \otimes_p 
\Theta^{\otimes_p (N-1)} \dket{V,\overline{V}}  \\
N_R M^\dagger & = & \dbra{W,\overline{W}}  
\Theta^{\otimes_p (N-1)} \otimes_p \Xi_1 \Omega_2^\star \dket{V,\overline{V}}  \\
M N_R^\dagger & = & \dbra{W,\overline{W}}  
\Theta^{\otimes_p (N-1)} \otimes_p \Omega_1 \Xi_2^\star \dket{V,\overline{V}} 
\eea
Hence
\be 
-i \comm{H_b}{MM^\dagger} = 
\dbra{W,\overline{W}} \Lambda \otimes_p \Theta^{\otimes_p (N-1)} \dket{V,\overline{V}} 
- \dbra{W,\overline{W}} \Theta^{\otimes_p (N-1)}\otimes_p \Lambda \dket{V,\overline{V}} 
\ee
with $\Lambda = i \left(\Omega_1 \Xi_2^\star - \Xi_1 \Omega_2^\star\right)$.
Expanding $\Lambda$ using \eref{Ealgebra} yields 
\be  
\Lambda = \sum_{\alpha\alpha'} E^{\alpha\alpha'} \otimes \Lambda^{\alpha\alpha'}
\ee 
with $\Lambda^{\alpha\alpha'}$ given by \eref{Laa}.

Next we consider the surface part of the unitary evolution. For the 
boundary
fields we obtain from Corollary \ref{bfcoro}
\bea 
-i \comm{b^L_1}{MM^\dagger} & = & \dbra{W,\overline{W}} 
\Gamma_L \otimes_p \Theta^{\otimes_p (N-1)} \dket{V,\overline{V}} \\
-i \comm{b^R_N}{MM^\dagger} & = & \dbra{W,\overline{W}} \Theta^{\otimes_p (N-1)}
\otimes_p \Gamma_R \dket{V,\overline{V}}
\eea
with 
\be 
\Gamma_B = -i \sum_{\beta} \sum_{\alpha\alpha'} 
\comm{b^L}{E^{\alpha\alpha'}} \otimes 
\Omega^{\alpha\beta}_1 \bar{\Omega}^{\alpha'\beta}_2, \quad B\in\{L,R\}.
\ee
With \eref{commE} this yields 
\be 
\Gamma_B = \sum_{\alpha\alpha'} E^{\alpha\alpha'} \otimes \Gamma_B^{\alpha\alpha'}
\ee
with $\Gamma_B^{\alpha\alpha'}$ defined by \eref{Gaa}.


Putting together the bulk and the surface contribution thus yields
\bea 
-i \comm{H}{MM^\dagger} & = & \dbra{W,\overline{W}} 
\left(\Gamma_L + \Lambda\right)\otimes_p \Theta^{\otimes_p (N-1)} 
\dket{V,\overline{V}} \nonumber\\
& & + \dbra{W,\overline{W}} \Theta^{\otimes_p (N-1)}
\otimes_p \left(\Gamma_R - \Lambda\right) \dket{V,\overline{V}}
\eea

For the dissipator part of the generator $\mathcal{L}$ \eref{LMM}
we have from Corollary \ref{disscoro}
\bea 
\mathcal{D}_1(MM^\dagger) & = &
\dbra{W,\overline{W}} \Delta_L \otimes \Theta^{\otimes (N-k)} \dket{V,\overline{V}} \\
\mathcal{D}_N(MM^\dagger) & = &  \dbra{W,\overline{W}} \Theta^{\otimes (N-1)}
\otimes \Delta_R \dket{V,\overline{V}}
\eea
with 
\be 
\Delta_{B} = \sum_{\beta} \sum_{\alpha\alpha'} 
\mathcal{D}^{B} \left(E^{\alpha\alpha'}\right) \otimes 
\Omega^{\alpha\beta}_1 \bar{\Omega}^{\alpha'\beta}_2, \quad B\in\{L,R\}.
\ee

Using \eref{DissE} one finds after relabeling of indices
\bea
\Delta_{B} 
& = & \sum_{\alpha\alpha'} E^{\alpha\alpha'} \otimes \Delta_B^{\alpha\alpha'}
\eea
with $\Delta_B^{\alpha\alpha'}$ defined by \eref{Daa}.
Thus
\bea 
\mathcal{L} (MM^\dagger) & = &
\dbra{W,\overline{W}} 
\left(\Delta_L + \Gamma_L + \Lambda\right)
\otimes_p \Theta^{\otimes_p (N-1)} \dket{V,\overline{V}} \nonumber \\
& & + \dbra{W,\overline{W}} \Theta^{\otimes_p (N-1)}
\otimes_p \left(\Delta_R + \Gamma_R - \Lambda\right) \dket{V,\overline{V}} = 0
\eea
by the Lindblad boundary matching condition \eref{LBMC}. \hfill {} \qed
\end{proof}

\section{The Heisenberg ferromagnet}

We have skirted the issue of existence of representations of the quadratic algebra
arising from the local divergence condition and vectors satisfying the Lindblad boundary
matching condition. In order to demonstrate that the matrix product construction
of the previous section is not only non-empty but also allows for concrete non-trivial
results we review the application to the isotropic Heisenberg ferromagnet 
\cite{Kare13a,Kare13b,Popk16}. Important other models where the matrix product construction
has been employed include the quantum XX-chain \cite{Znid10},
one-dimensional Hubbard model \cite{Pros14}
and the spin-1 Lai-Sutherland chain \cite{Ilie14}.

\subsection{Definitions and notation}
\label{Sec:Definotat2}

It is expedient to introduce
the Levi-Civita symbol 
$\varepsilon_{\alpha\beta\gamma}$ (defined for $\alpha, \beta, \gamma
\in \{1,2,3\}$) by
$\varepsilon_{123} = 1$ and 
$\varepsilon_{\alpha\beta\gamma}=(-1)^\pi\varepsilon_{\pi(\alpha\beta\gamma)}$ 
for any permutation $\pi(\cdot)$. We also define $\zeta_{0\alpha\beta} = \zeta_{\alpha 0\beta} = \zeta_{\alpha\beta 0} 
= \delta_{\alpha,\beta}$ for
$\alpha, \beta \in \{0,1,2,3\}$ and $\zeta_{\alpha\beta\gamma} = i
\epsilon_{\alpha\beta\gamma}$ for
$\alpha, \beta, \gamma \in \{1,2,3\}$ and introduce
the two-dimensional unit matrix and the Pauli matrices
\bel{Pauli}
\sigma^0 \equiv \mathds{1} := \left( \ba{cc} 1 & 0 \\ 0 & 1 \ea \right), \,
\sigma^1 := \left( \ba{cc} 0 & 1 \\ 1 & 0 \ea \right), \,
\sigma^2 := \left( \ba{cc} 0 & -i \\ i & 0 \ea \right), \,
\sigma^3 := \left( \ba{cc} 1 & 0 \\ 0 & -1 \ea \right)
\ee
which form a complete basis of $\mathfrak{End}(\C^2)$. They satisfy
\bel{Paulirelations}
\sigma^{\alpha}\sigma^{\beta} 
= \sum_{\gamma=0}^3 \zeta_{\alpha\beta\gamma} \sigma^\gamma
\ee
For $\alpha\in\{1,2,3\}$ the matrices $\sigma^\alpha_k$ are related by a
unitary transformation $U$ with the property 
\bel{Udef}
U\sigma_k^\alpha U^\dagger = \sigma_k^{\alpha+1}, 
\quad \forall k \in \{1,\dots,N\}, \quad
\alpha \ \mathrm{mod} \ 3.
\ee
Straightforward computation shows that this transformation is realized by
the tensor product
\bel{U}
U = u^{\otimes N}
\ee
with
\be 
u = \frac{1}{\sqrt{2}} \left( \ba{cc} 1 & -i \\ 1 & i \ea \right)
\ee
which is unique up to a non-zero factor.

We shall also use the notation $\hat{n} \equiv E^{00} = (1+\sigma^z)/2$, 
$\sigma^+ \equiv E^{01} = (\sigma^x + i \sigma^y)/2$, 
$\sigma^- \equiv E^{10} = (\sigma^x - i \sigma^y)/2$, 
$\hat{v} \equiv E^{11} = (1-\sigma^z)/2$ 
and the representation of the local basis vectors as column vectors as
\be
|0) := \left( \ba{c} 1 \\ 0 \ea \right) , \quad |1) := \left( \ba{c} 0 \\ 1 \ea \right).
\ee
For later use we
also introduce the notation $\sigma^x \equiv \sigma^1$, $\sigma^y \equiv \sigma^2$, 
$\sigma^z \equiv \sigma^3$ and
the three-vectors $\vec{\sigma} = (\sigma^1,\sigma^2,\sigma^3)$
with the dot product 
$\vec{A} \cdot \vec{B} := \sum_{i=1}^3 A^i B^i$. Here the
$A_i$ and $B_i$ can be real numbers or Pauli matrices. The reason for introducing
this definition is the interpretation of the upper indices of the
Pauli matrices as the components of the (quantum) angular momentum
vector of an atom in the coordinate directions $x,y,z$ of $\R^3$. If $\vec{A} \in \R^3$ 
is a vector of (Euclidean) length $\vec{A} \cdot \vec{A}=1$, then the 
quantum expectation $\exval{\vec{A} \cdot \vec{\sigma}}$ is the mean of the
projection of the angular momentum vector in the direction defined by the 
vector $\vec{A}$.

The Lie algebra $\mathfrak{gl}_2(\C)$ 
with generators $X^\alpha$, $\alpha \in \{0,1,2,3\}$ is
defined by Lie brackets
\bea
\label{gl2def1}
\comm{X^0}{X^\alpha} & = & 0 \\
\label{gl2def2}
\comm{X^\alpha}{X^\beta} & = & 2 i \sum_{\gamma=1}^{3}
\varepsilon_{\alpha\beta\gamma} X^\gamma, \quad
\alpha, \beta \in \{1,2,3\} .
\eea
The two-dimensional unit matrix $\mathds{1}$ and Pauli matrices $\sigma^\alpha$ 
\eref{Pauli} are representation matrices for $\mathfrak{gl}_2(\C)$ with the
Lie-bracket represented by the commutator. 
Since 
$\sigma^\alpha_k \sigma^\alpha_l = \sigma^\alpha_l \sigma^\alpha_k$ for
$l\neq k$ it follows that also $\mathbf{1} \in \mathfrak{P}$ together with
\bel{SU2}
S^{\alpha} = \sum_{k=1}^{N} \sigma^\alpha_k \in \mathfrak{P}.
\ee
are representation matrices of $\mathfrak{gl}_2(\C)$. 
We say that an endomorphism $G$ on
$\mathfrak{P}$ is $SU(2)$-symmetric if its representation matrix satisfies
$\comm{G}{S^\alpha}=0$ for
$\alpha\in\{1,2,3\}$.

We also define the generators
\bel{Spmdef}
X^\pm := \half \left( X^1 \pm i X^2 \right), \quad X^z := \half X^3.
\ee
In terms of these generators the defining relations \eref{gl2def1}, \eref{gl2def2} 
of $\mathfrak{gl}_2(\C)$ read
\bea
\label{gl2alt1}
\comm{X^0}{X^{\pm,z}} & = & 0 \\
\label{gl2alt2}
\comm{X^+}{X^-} & = & 2 X^z, \quad \comm{X^z}{X^\pm} = \pm X^\pm .
\eea
An infinite-dimensional family of representations $X^0 \mapsto I$, 
$X^{\pm,z} \mapsto  S^{\pm,z}$
is given by matrices $I,S^{\pm,z}$ with matrix elements
\bel{gl2repinf}
I_{kl} = \delta_{k,l}, \quad
S^+_{kl} = l \delta_{k+1,l}, \quad S^-_{kl} = (2p-l) \delta_{k,l+1}, \quad 
S^z_{kl} = (p-l) \delta_{k,l}
\ee
for the non-negative integers $k,l \in \N_0$ and parameter $p\in\C$.

\subsection{Boundary-driven Lindblad-Heisenberg chain}
\label{Sec:LH}

We consider an open chain of $N \geq 2$ quantum spins in contact with boundary
reservoirs for which we wish to construct the stationary density matrix
defined by \eref{NESSQHdef}.
For the  unitary part of the time evolution we consider
the isotropic spin-1/2 Heisenberg Hamiltonian \cite{Heis28,Baxt82} defined 
with the dot-product by
\begin{equation}
\label{Hamiltonian} 
H = \sum_{k=1}^{N-1} \vec{\sigma}_k \cdot \vec{\sigma}_{k+1}.
\end{equation}
for $N$ quantum spins at positions $k$ along the chain.

Before defining the boundary dissipators we point out that $H$ is manifestly
rotation invariant in $\R^3$ which due to the quantum nature of the spin is equivalent
to the symmetry $[H,S^{\alpha}]=0$ under the Lie-algebra $SU(2)$ with representation
matrices \eref{SU2}.
Thus the spin components are locally conserved with associated
locally conserved currents $j_k^\alpha$ defined by \eref{Lindad} with
$F = \sigma^\alpha_k$. For $1 < k < N$ the action of the adjoint generator
\eref{Lindad} yields
\be 
\mathcal{L}^\dagger (\sigma^\alpha_k) = j_{k-1}^\alpha - j_k^\alpha
\ee
with
\bel{localcurr}
j_k^\alpha = 2 
\sum_{\beta=1}^{3} \sum_{\gamma=1}^{3}
\varepsilon_{\alpha\beta\gamma} \sigma_{k}^{\beta}\sigma_{k+1}^{\gamma},
\quad
1 \leq k < N.
\ee
In the steady
state the current expectations $j^{\alpha} := \langle j_{k}^{\alpha}\rangle$ 
are position-independent.

We choose two boundary Lindblad operators $D^{L,R}$ to favour a relaxation of the
boundary spins towards target states given by 
density matrices $\rho_{L},\rho_{R}$ 
satisfying $\mathcal{D}_{1}(\rho_{L})= \mathcal{D}_{N}(\rho_{R})=0$. As target
states we choose fully polarized states of one boundary spin 
\be 
\rho_{L}=\frac{1}{2}\left(\mathds{1} + 
\vec{n}_{L} \cdot \vec{\sigma} \right) \otimes \tilde{\rho} , \quad
\rho_{R}=\tilde{\rho} \otimes \frac{1}{2}\left(\mathds{1} + 
\vec{n}_{R} \cdot \vec{\sigma} \right) 
\ee 
where $\left\vert \vec{n}_{L}\right\vert
=\left\vert \vec{n}_{R}\right\vert =1$ and $\tilde{\rho}$
is an arbitrary reduced density matrix for the remaining $N-1$ spins. 
The reduced single-site boundary density matrix 
$\rho^{(1)}_B = \frac{1}{2}\left(\mathds{1} + 
\vec{n}_B \cdot \vec{\sigma} \right)$
is a pure state since for a projection direction given by 
\be 
\vec{n}_B
= (\sin(\phi_B)\cos(\theta_B),\sin(\phi_B)\sin(\theta_B),\cos(\phi_B)).
\ee 
One has
$\rho^{(1)}_B = | \psi_B ) \otimes ( \psi_B |$ with
\be 
| \psi_B ) = \rme^{i\alpha_B} \left( \ba{c} \cos{(\phi_B/2)}\rme^{-i\theta_B/2} \\
\sin{(\phi_B/2)}\rme^{i\theta_B/2} \ea \right)
\ee 
and arbitrary phase $\alpha_B \in [0,2\pi)$. The notion ``full polarization''
means that the expectation of the spin projection $\vec{n}_B \cdot \vec{\sigma}$ 
in the space-direction
defined by $\vec{n}_B$ is given by $\exval{\vec{n}_B \cdot \vec{\sigma}} = 1$.

Due to the rotational symmetry \eref{SU2} of $H$ only the 
angle between the two boundary polarization vectors plays a role.
Therefore we may, without loss of generality, choose $\phi_L=\phi_R=\pi/2$ and
fix the coordinate frame in $\R^3$ such that the $X$--axis points in the 
$\vec{n}_{L}$ direction (corresponding to $\theta_L = 0$)
and to let 
the $XY$-plane be spanned by the family vectors $\vec{n}_{R}(\theta)$, i.e.,
\bel{boundaryspins}
\vec{n}_{L} = (1,0,0), \quad
\vec{n}_{R} = (\cos\theta,\sin\theta,0) , \quad 0 \leq\theta \leq \pi 
\ee
corresponding to $\theta_R=\theta$. 

It is easy to verify that there are two families of local Lindblad 
operators satisfying $\mathcal{D}_{1}[\rho_{L}]=0$, viz.
$D_1^R = a(\sigma_1^{2}+i\sigma_1^{3})+b(\mathbf{1}-\sigma_1^{1})$
and $D_1^{R'} = a' \mathbf{1} + b'\sigma_1^{1}$. Following \cite{Kare13a,Kare13b}
we choose $D^R$ with $b=0$ (so that 
$\Tr (D_1^R) = 0$) and coupling strength $a=\sqrt{\Gamma}$. Similarly, we choose for the
right boundary site $N$ the rotated projection to arrive at
\begin{equation}
\label{XLXR}
D_1^R=\sqrt{\Gamma}(\sigma_1^{2}+i\sigma_1^{3}), \quad D_N^L=\sqrt{\Gamma}
(\sigma_N^{2}\cos\theta-\sigma_N^{1}\sin\theta+i\sigma_N^{3}).
\end{equation}
Then in absence of the
unitary term in \eref{NESSQHdef} the boundary spins relax with characteristic
times $\propto \Gamma^{-1}$ to approach $\rho_{L},\rho_{R}$: 
Writing $\rho_{L}(t) = 1/2 (\sigma_1^0 + x(t) \sigma_1^1 + y(t)
\sigma_1^2 + z(t) \sigma_1^3)\otimes\tilde{\rho}$ one has 
$x(t) = 1 + (x(0)-1)\exp{(-4\Gamma t)}$, $y(t) = y(0) \exp{(-2\Gamma t)}$,
$z(t) = z(0) \exp{(-2\Gamma t)}$, 
and similarly for $\rho_{R}(t)$.

\begin{rem}
In the untwisted case $\vec{n}_L=\vec{n}_R:=\vec{n}$
corresponding to $\theta=0$ the Lindblad equation \eref{NESSQHdef} for the stationary 
density matrix is trivially solved by \cite{Pros11}
\be 
\rho_N(\Gamma,0) = 
\left(\frac{\mathds{1} + 
\vec{n} \cdot \vec{\sigma}}{2} \right)^{\otimes N}.
\ee
This is a pure state of the form
$\rho_N(\Gamma,0) = \ket{\Psi}\bra{\Psi}$
where
$\ket{\Psi} = \ket{\psi}^{\otimes N}$
and
\be 
\ket{\psi} = \frac{1}{\sqrt{2}} \left(\ba{c} 1 \\ 1 \ea \right).
\ee
This pure state is not of the form $\exp{(-\beta H)}/Z$ for any
$\beta$ and therefore not an equilibrium state.
\end{rem}

\subsection{Matrix product solution}

From now on we exclude $\theta=0$ so that the boundary
coupling introduces a twist in the $XY$-plane, which
drives the system perpetually out of equilibrium.

\begin{theo}
\label{Theo2}
Let $I, S^{\pm,z} \in \mathfrak{End}(\mathfrak{A})$ be the infinite 
dimensional representation \eref{gl2repinf} of $\mathfrak{gl}_2(\C)$
with 
representation parameter
\be 
p = i \Gamma^{-1}
\ee
and  let
$\Omega^{00} = - \Omega^{11} = i S^{z}$, $\Omega^{01}=i S^+$,
$\Omega^{10}=i S^-$. Furthermore, let
\be
\bra{\overline{W},W} = \bra{0}\otimes\bra{0}, \quad 
\ket{V,\overline{V}} = \sum_{m,n=0}^\infty \left(-\cot{\frac{\theta}{2}}\right)^{m+n}
{2p \choose m} {2\bar{p} \choose n} \ket{m}\otimes\ket{n}.
\ee
Then for $\rho$ in matrix product from \eref{rhoxi} and $U$ defined by \eref{Udef}
the density matrix
\bel{rhoNESS} 
\rho_N(\Gamma,\theta)= U \rho U^\dagger
\ee
is the unique solution of the quantum master equation \eref{NESSQHdef} for the
Heisenberg ferromagnet \eref{Hamiltonian} with boundary dissipators \eref{XLXR}. 
\end{theo}

Uniqueness is guaranteed by the structure of the
Lindblad dissipators, see \cite{Pros15}. The proof of \eref{rhoNESS}
follows from verifying 
the local divergence
condition \eref{LDC} with $\Xi^{00} = \Xi^{11} = i I$, $\Xi^{01} = \Xi^{10} = 0$
and the Lindblad boundary matching condition \eref{LBMC}
by (somewhat lengthy but straightforward) explicit computation \cite{Kare13a,Kare13b}.
We summarize the main conclusions of \cite{Kare13a,Kare13b,Popk16} drawn from
Theorem \ref{Theo2} and the underlying quadratic algebra and Lindblad boundary matching
property.\\

\noindent (1) Proposition \ref{Zprop} yields for the non-equilibrium
partition function \eref{partfundef}
\be 
Z_N(\Gamma,\theta) = \bra{\overline{W},W}\Theta_0^N\ket{V,\overline{V}} 
\ee
with $\Theta_0 = 2 S^z_1 S^z_2 + S^+_1 S^+_2 + S^-_1 S^-_2$ defined by \eref{Theta0def}.
Dropping the arguments $\Gamma,\theta$, 
the stationary magnetization currents are then given by
\bel{currents}
j^x_N = -8ip \frac{Z_{N-1}}{Z_N}, \quad
j^y_N = - \cot{\frac{\theta}{2}} j^x_N, \quad
j^z_N = - 4 \frac{\frac{\rmd}{\rmd \theta}Z_{N-1}}{Z_N}.
\ee
Based on numerically exact computation up to $N=100$ we conjectured that
for any fixed
coupling strength $\Gamma$ and any fixed $0 < \theta < \pi$ one has  \cite{Kare13b}
\bel{Zconj} 
\lim_{N\to\infty} N^2 \frac{Z_{N-1}(\Gamma,\theta)}{Z_N(\Gamma,\theta)}
= \frac{1}{4} \theta^2.
\ee
For the currents this result implies
\be 
\lim_{N\to\infty} N^2 j^x_N(\Gamma,\theta) = \frac{2\theta^2}{\Gamma}, \quad
\lim_{N\to\infty} N j^z_N(\Gamma,\theta) = 2\theta .
\ee 

Some rigorous results have been obtained for
the Zeno limit $\Gamma\to\infty$ \cite{Popk16}.
Rescaling the normalization factor yields a finite limit
\bel{YGt}
\tilde{Z}_N(\theta) := \frac{1}{4} \lim_{\Gamma\to\infty} \Gamma^2 Z_N(\Gamma,\theta)
\ee
which was computed explicitly.
Then for small twist angle $\theta = o(1/N)$ the conjecture \eref{Zconj}
can be proved rigorously.  For the currents one therefore finds
\begin{theo}
Let $\tilde{j}^{\alpha}_N(\theta) := \lim_{\Gamma\to\infty} 
j^{\alpha}_N(\Gamma,\theta)$ be the stationary
currents of the Heisenberg chain in the Zeno limit. Then for any $N$ one has
\be 
\tilde{j}^{x}_N(\theta) = \tilde{j}^{y}_N(\theta) = 0 
\quad \forall \theta \in [0,\pi[
\ee
and for any real $\epsilon > 0$ and real $\theta_0 > 0$
\be 
\lim_{N\to\infty} N^{2+\epsilon} \tilde{j}^{z}_N\left(\frac{\theta_0}{N^{1+\epsilon}}\right) = 2 \theta_0.
\ee
\end{theo}

The first statement is a trivial consequence of the explicit expressions
\eref{currents} and the result that $\tilde{Z}_N(\theta)$ is finite and
non-zero. The second statement follows from the explicit form of 
$\tilde{Z}_N(\theta)$ given in \cite{Popk16}.

\noindent (2) In terms of
\be 
B^x := \Theta^{01}+\Theta^{10}, \quad
B^y := i \left(\Theta^{01}-\Theta^{10}\right), \quad B^z = \Theta^{00}-\Theta^{11}
\ee
the multiplication property \eref{Paulirelations} yields
\be 
\exval{\sigma_k^\alpha}_N = \frac{S_{k,N}^\alpha(\Gamma,\theta)}{Z_N(\Gamma,\theta)}
\ee
with
\be 
S_{k,N}^\alpha(\Gamma,\theta) = \bra{\overline{W},W}\Theta_0^{k-1}B^\alpha
\Theta_0^{N-k} \ket{V,\overline{V}}.
\ee

\noindent (3) The quadratic algebra implies that the 
operators $\Theta_0$ and $B^\alpha$ satisfy the 
remarkable {\it cubic} relation
\bel{cubicB}
\comm{\Theta_0}{\comm{\Theta_0}{B^\alpha}} + 2 \{\Theta_0,B^\alpha\} - 8 p^2 B^\alpha = 0 
\ee
which was obtained earlier on the basis of computer algebra \cite{Pros11}.
This relation induces recursion relations for the unnormalized correlation functions 
$Z_N \exval{\sigma_{k_1}^{\alpha_1} \dots \sigma_{k_n}^{\alpha_n}}$. In particular,
with
\bea 
& & B^x = S^+_1 S^+_2 - S^-_1 S^-_2, \quad
B^y = S^z_1 (S^-_2 - S^+_2) + (S^-_1 - S^+_1) S^z_2, \nonumber \\
& & B^z = i S^z_1 (S^-_2 + S^+_2) - i (S^-_1 + S^+_1) S^z_2.
\eea
one finds for the unnormalized one-point function (dropping the arguments)
\be 
S_{k+2,N+1}^\alpha + S_{k,N+1}^\alpha -2 S_{k,N}^\alpha 
+ 2(S_{k,N}^\alpha + S_{k+1,N}^\alpha) - 8 p^2 S_{k,N-1}^\alpha = 0.
\ee
By setting $r=k/N$ and taking the continuum limit $k,N\to\infty$ such that 
the macroscopic coordinate $r$ remains fixed this recursion together with
\eref{Zconj} yields the simple ordinary differential equation 
$m''(r) + \theta^2 m(r) = 0$ for the large-scale magnetization profile
$m^\alpha(r) := \lim_{k,N\to\infty} \exval{\sigma_k^\alpha}_N$.
The boundary conditions are given by the microscopic complete polarizations so that
\be 
m^x(r) = \cos{(\theta r)}, \quad m^y(r) = \sin{(\theta r)}, \quad m^z(r) = 0, \quad
0 \leq r \leq 1 
\ee
corresponding to a spin helix state \cite{Popk16}.
It was also pointed out in \cite{Popk16} that this implies a strongly sub-diffusive 
current $0 = \lim_{N\to\infty} N j^x_N = \lim_{N\to\infty} N j^y_N$ inside the
twist-plane, and a ballistic current perpendicular to it if one
regards the associated magnetization gradients as analogous to
density gradients in classical transport. 

Correlation functions were computed in \cite{Buca16} using the cubic relation
\eref{cubicB} and the resulting continuum approximation. Remarkably they are
of a form similar to what was obtained for the symmetric simple
exclusion process with open boundaries, using the fluctuating hydrodynamics
approach \cite{Spoh83}. This similarity suggest that also the boundary
driven quantum problem may be understood in terms of fluctuating hydrodynamics.

\section*{Acknowledgements}

VP and GMS thank T. Prosen for useful discussions and DFG for financial support.

\end{document}